# Network Selection schemes in Heterogeneous Wireless Networks

Fayssal Bendaoud[1*], Marwen Abdennebi[2#], Fedoua Didi[1*]
[1]Laboratory of Telecommunication Tlemcen (LTT)
[*]Téchnology's Faculty, Tlemcen's University
[2]Laboratory of Processing and Transmission of Information (L2TI)
[#]Physics Department, University of North Paris

*Abstract*—**Heterogeneous Wireless Networks HWNs are a combined networks made of different Radio Access Technologies RAT. Next Generation Networks NGN or HWNs will provide a high bandwidth connectivity and high data throughput with a smooth support for the user's QoS requirements, in this context, users with multi-interface terminals will be able to connect to different wireless technologies such as: 802.16, 802.11 families and cellular families UMTS, HSPA and LTE in the same time.
The idea of NGN or HWNs is that users will not be tied with a contract with one single operator but, users will be able to choose the Radio Access Network RAT considering the user's QoS requested. This paper focus on the network selection strategies and the inter technologies Handoff, we will present a descriptive of the existed methods of network selection, we will discuss the merits and the weakness of such method and we will give ours point of view.**

I. INTRODUCTION

One of the most important features of the Next-Generation Network is the heterogeneous wireless access in which a mobile device can choose the most suitable radio access technology according to its needs i.e. the QoS requirements.

A HWN is a set of overlapped RATs such as cellular networks: GSM, UMTS, HSPA and LTE, WLAN: 802.11.x and the WMAN: 802.16 through a common platform. Modern wireless terminals are multi-interfaces, these multi-interfaces terminals can roam among HWN with a seamless mobility and service contiguity, so users can enjoy multiple wireless services with a smooth manner.

For user's point of view, the problem is selecting "best" radio access at given time. To choose the network to which to connect, we must start with discovering the networks. Networks discovery means that terminal should find and determine the available RAT within an area. [1]

An important challenge for the HWN architecture is the Radio Resource Management RRM. The current RRM solutions consider only the case of single RAT i.e. each technology has its own RRM strategy implemented. The RRM functionalities and mechanisms are among others: network selection, bandwidth allocation, congestion control, power control, handover control and admission control etc…None of the existed RRM strategies are suitable for the heterogeneous environments because they consider only the situation with one RAT. The solution then is the use of Common RRM CRRM or the Joint RRM JRRM. The CRRM has been proposed to manage the radio resource utilization among a set of RATs in a seamless way. Its concept is based on two-tier model CRRM and RRM entities. The RRM entity manage the radio resource unit in a single RAT, and the CRRM controls the RRM entities and communicate with other CRRM, figure1. [14]

This paper focus on the network selection strategies and mechanisms, we will present and discuss the proposed solutions in the literature. An important concept should be considered is the handover. The handover process means that the handover process enables a mobile terminal to maintain the call in progress, during a movement, which causes the mobile to change cell. Indeed, when the signal transmission between a terminal and a base station weakens, the mobile phone software looks for another base station available in another cell that is capable of ensuring the new communication the best conditions; it is based on the network selection idea. It exist several type of handover, horizontal HO and vertical HO, we will explain it in the next section.

The rest of this paper is organized as follows, in the section two we will present the handover process, in section three the related works and a discussion will be offered, and conclusion with perspectives in the section four.

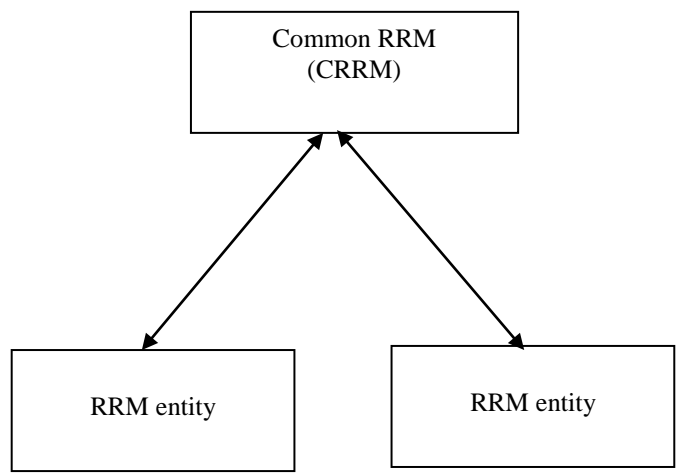

Figure 1- The CRRM/RRMs model [14]

## II. THE HANDOVER PROCESS CHARACHTERISTICS

We can classify the Handover process in accordance with the following characteristics. We have handover between radio channels of the same cell called intracell handover, the horizontal handover between different cell of the same network, and we have the vertical handover between different networks.

The handover can be Network Controlled/Mobile Assisted (NCHO/MAHO) when a network entity is responsible for monitoring and guiding the handover, exploiting information and measurements gathered form the mobile terminal, it can be Mobile Controlled/Network Assisted (MCHO/NAHO) when the mobile terminal has the principal control over the handover exploiting information provided by the network. For the approach of NCHO/MAHO, as the decision point is located in the network, the overall network load is better balanced, due to exploiting the more exactly knowledge of the network's conditions. On the other hand, in MCHO/NAHO approach, the decision depends on local conditions and metrics measured by the mobile terminal. This facilitates and improves the performance of the handover initiation decision. The upwards handover occurs between a network supporting high data rate and small coverage or a network achieving higher coverage but lower data rate. The opposite is the downwards handovers when the mobile node moves from a large network cell with a low data rate to a small network cell which supports high data rates.

The handover is soft/smooth when the mobile terminals build a connection to a point of attachment before to the release of the previous attachment point. The mobile terminal was listening to a set of candidate attachment points (access points) at the same time before selecting one of them. It is also referred to as make-before break handover and is an essential to achieve a seamless made after the release of the old one; the handover is called hard or break-before-make handover. In such a case, the mobile terminal is able to communicate each time with only one access point. Another characteristic in the components that are involved in the overall handover process. Components should be sensitive of the existence and adaptable to various wireless networking technologies, should also be capable of handling more complex and dynamic situations and able to manage user's mobility, maintaining seamless service provisioning and d) have low power requirements. [13]

Considering all this characteristics, handover requires multimode user devices, either using multiple radio interfaces adapting to different wireless networks, while they should be provided with a friendly graphical user interface to specify & alter user preferences, requirements and constraints in an easy manner.

## III. RELATED WORKS

### A. Heteregeneous wireless networks

Heterogeneous wireless system is composed of several wireless technologies, which connect user terminals to the internet through a common core network or a backbone networks. The wireless access technologies overlap and result in a HWN combined several different access technologies.

With the emergence of mobile terminals and smartphones, user have the choice to connect to different wireless technologies with the multi interfaces terminals. The important property of HWN is the capacity of providing the best feature of each network like low cost and simplicity and high speed of Wlans, the wide range of mobility of the cellular networks and the wider coverage of the satellite networks. So, by using a HWN we can profit from the best QoS offered by a heterogeneous system. This characteristics and features mentioned above cannot be obtained by a homogeneous environment.Figure2

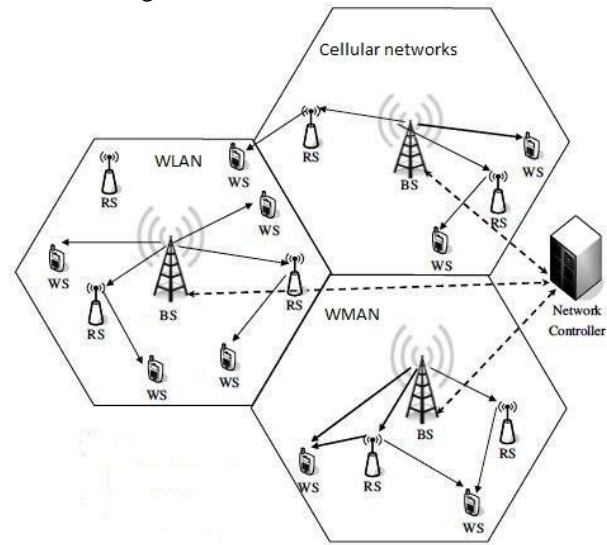

Figure 2- HWN system

### B. Network selection: a state of the art

In this sub-section, we will present and discuss the most recent a representative method for the network selection and radio resource management in a heterogeneous wireless system. They are mainly divided in three category, network centric approach, user centric approach and the collaborative one.

#### 1) Network controlled scheme

In this approach, the most important concern is dealing with the optmisation problem of bandwidth allocation by the network side; the decisoons are made in the network side and they consider only network profit. The methods using this approach are presented below.

##### a) Stochastic programing

Stochastic programs are mathematical programs where some of the data incorporated into the objective or constraints are uncertain. Uncertainty is usually characterized by a probability distribution on the parameters. Although the uncertainty is rigorously defined, in practice it can range in detail from a few scenarios (possible outcomes of the data) to specific and precise joint probability distributions. The outcomes are generally described in terms of elements w of a set W. W is the set of possible demands. When some of the data is random, then solutions and the optimal objective value of the optimization problem are themselves random. Ideally, we would like one decision and one optimal objective value. Here the goal is to find a solution which is feasible for all such data and optimal in some sense.

The authors of [2] have deployed a mathematical programming technique used in decision-making problems under uncertainty called stochastic programming (SP). With the probabilistic nature of requests in wireless heterogeneous networks, the scheme actually uses a subset of SP called stochastic linear programming (SLP). For example, cellular network and satellite networks provide a data service with fixed bandwidth requirement. The idea is to associate probabilistic demands with fixed significant probability, then formulate given scenario with allocation, underutilization, and rejection along with the fixed probability. Thus, the goal is to obtain maximum allocation in both networks while minimizing cost of resource underutilization and demand rejection.

This method is the first mathematical attempt that addresses joint resource management, in which several access networks in HWNs provide bandwidth to users. However, the scheme support single common service with fixed required bandwidth, which is not suitable to variety of services having various bandwidth requirements in networking today.

*b) Game theory*

Game theory is an area of applied mathematics, which attempts to capture behavior in strategic situations, in which an individual's success in making choices depends on the other's choices. With this special type of N-person cooperative game, each network cooperates to provide the requested bandwidth to a new connection using coalition form and characteristic function.

Game theory is relevant to parlor games such as poker or bridge, most research in game theory focuses on how groups of people interact. There are two main branches of game theory: cooperative and noncooperative game theory. Noncooperative game theory deals largely with how intelligent individuals interact with one another in an effort to achieve their own goals. A game is cooperative if the players are able to form binding commitments. For instance, the legal system requires them to adhere to their promises. In noncooperative games, this is not possible.

Authors presents works that consider the radio resource management in a single RAT, such as in [3], the control admission problem was formulated as a non-cooperative game. The formulation considered the choice of a user to churn from a current provider to another. The decision on whether a new user can be admitted or not and the allocated transmission are determined from the Nash equilibrium.

In [15], authors propose a cooperative game structure to solve the bandwidth allocation problem in a HWN. The problem is formulated as a cooperative game and the solution is the amount of bandwidth available to each connection is obtained from the Shapley value. The difference between the cooperative and non-cooperative approaches lies in the fact that the first one is a group-oriented, while the second is individual oriented method. In the non-cooperative approach, each network is interested only by its benefices to maximize its profits, it is egoist and selfish i.e. in a non-cooperative environment, all networks are competing with each other to achieve their own objectives. However, in a cooperative approach, groups of players seeking an equitable distribution of resources.

*c) Degradation utility*

The objective of this method is dealing with the different user priorities; by degrading lower priority users, higher priority users can get more bandwidth (the released bandwidth from the degraded users). Therefore, for each service, the network operator specifies the quantity of bandwidth needed for the different class of QoS (excellent, good, basic and rejected). After that, a table of reward is defined, for each QoS class associated with a service application (video, VOIP and data), a reward is defined. The degradation utility is the ratio of released bandwidth and the lost rewards points (released bandwidth is the difference of bandwidths before and after degradation and the lost reward points is difference of reward point before and after degradation).

When a new connection arrive, network operator finds all possible degradable connections, computes their degradation utilities, and begins by degrading the connection that gives highest utility. [4]

*2) User-controlled scheme*

In this approach, decisions are made a user's side, they consider only user profit, and they do not care about network load balancing. The most concern of this approach is the network selection issue, which mean finding the most appropriate network. The big problem of this methodology is the network congestion and this result in a degradation of the ongoing users. The methods using this approach are presented below.

*a) Analytical hierarchy process*

The analytic hierarchy process (AHP) is a structured technique for organizing and analyzing complex decisions, based on mathematics. It has particular application in group decision making and is used around the world in a wide variety of decision situations.

An AHP is employed for decision problems to weigh QoS factors. An order preference technique based on Grey Relational Analysis (GRA) is then applied to rank the evaluation alternatives. AHP is used to solve complex decision-making problems involving different areas, including planning, resources assessment, resource allocation and policy selection. After that, GRA is direction among the many applications of grey system theory that solves the complicated interrelationships among multiple performance characteristics by optimizing grey relational ranks. [6]

The authors of [5] propose AHP method to weigh QoS factors and using GRA to rank the networks to solve network selection problem. The authors separate between QoS parameters, we have network conditions parameters and user preferences parameters. The hierarchy is as follows: each kind of QoS is placed at a level, at the top, we have the objective i.e. QoS, in the second level, we have the network conditions factors as: throughput, cost, reliability, security and timeliness. At the third level, they decompose the factors of the second level into sub factors. In the last level, the available solutions are placed.

After that, the algorithm collect the user's data by AHP to get the global factors of second level and local weights of third level factors and then the final weights are computed. Meanwhile, networks data are controlled by GRA, and the perfect network performance is defined by calculation of the Grey Relational Coefficient (GRC), which gives grey relationship between best network and the other. Then, the network with the biggest GRC is more desirable.

### b) Consumer surplus

An economic measure of consumer satisfaction, which is calculated by analyzing the difference between what consumers are willing to pay for a good or service relative to its market price. A consumer surplus occurs when the consumer is willing to pay more for a given product than the current market price. The consumer surplus is the amount that consumers benefit by purchasing a product for a price that is less than they would be willing to pay.

Authors in [7] propose a costumer surplus. The scheme is proposed for a non-real time services. The algorithm is as follows: The users survey the radio interface and determine a list of available access networks. Next, they calculate transfer amounts of networks on the list by taking its average of the last five data transfers and then get the execution times. After that, the users compute predicted utility i.e. the relationship between the budget and the user's flexibility in the transfer completion time. Finally, for each candidate network; the users compute consumer surplus, which is the difference between utility and cost charged by the network, and they choose the best one to request for connection.

### c) Profit function

A profit function is defined in economics as: $\pi(p, w) = \max_x pf(x) - wx$, w and x are two vectors of factors prices and p is the output price.

The authors in [8] consider a different definition to handle handoff selection in HWNs. They calculate profit for each handover by a target function with two parameters: bandwidth gain and handoff cost. Furthermore, they classified handoffs into reactive and proactive handoff. A reactive handoff is initiated whenever a mobile equipment roam out of its current cell, while proactive handoff is initiated at periodical discrete period when connection experience can be improved.

The gain is calculated using the following parameters: the maximum bandwidth provided by a RAT for a single user and the percentage of capacity used, bandwidth requirement of service needed and the bandwidth.

The authors define a cost that's represent the data, which could be transmitted during handoff process, it correspond to the data quantity lost due to handoff delay. The profit then is, the difference between cost and gain, and at before each handoff operation, the user terminal compare profit from different RAT and choose on that give biggest profit.

### 3) Combined scheme

This approach try to balance between the network conditions and the users profits. Besides the two described methods, it's the most appropriate. The methods using this approach are presented below.

### a) Fuzzy logic

Fuzzy logic is a mathematical logic concept that attempts to solve problems by assigning values to an imprecise spectrum of data in order to arrive at the most accurate conclusion possible. Fuzzy logic is designed to solve problems in the same way that humans do: by considering all available information and making the best possible decision given the input. [9]

Authors in [5] develop an algorithm based on fuzzy logic controller (FLC) and calculate the robustness ranking of the list of possible selected networks. They start by dividing decision making into three steps: pre-selection, discovery, and make decision. Pre-selection phase eliminate inappropriate access networks from further selection, it takes its parameters from user, application, and network to. If no RAT are suitable for user's requirement, the system demand user to reduce its requirements. The second step is the discovery phase; the authors fuzzify the values of the variables (network data rate, SNR, and application's data rate requirement) into grade of membership in fuzzy set. The discovery phase role is to power-up users when no current connections exist, and when a connection is already established but QoS is not meeting the demands at the same time other potential networks become available. The results are used as input in the final stage; finally, overall ranking is obtained through defuzzification with weighted average method. The current scheme deployed it for network selection; FLC gives a good result in this case of few metrics. However, if the number of metrics increases, the system may become very complex and may give erroneous results.

### b) TOPSIS

The Technique for Order of Preference by Similarity to Ideal Solution (TOPSIS) is a multi-criteria decision analysis method; the TOPSIS is based on the concept that the chosen alternative should have the shortest geometric distance from the positive ideal solution and the longest geometric distance from the negative ideal solution. It is a method of compensatory aggregation, which compares a set of alternatives by identifying weights for each criterion, normalizing scores for each criterion and calculating the geometric distance between each alternative and the ideal alternative, which is the best score in each criterion. An assumption of TOPSIS is that the criteria are monotonically increasing or decreasing. Normalization is usually required as the parameters or criteria are often of incongruous dimensions in multi-criteria problems. The Compensatory methods such as TOPSIS allow trade-offs between criteria, where a poor result in one criterion can be neglated by a good result in another criterion. This provides a more realistic form of modelling than non-compensatory methods. [10]

Without deep details, the authors of [11] proposed an algorithm for path selection on multi-homed end-hosts based on TOPSIS. In this mechanism, the authors collected parameters from both network level (QoS parameters: bandwidth, delay, jitter, and BER) and application level (traffic class: conversational, streaming, interactive, and background). They deployed TOPSIS for their Score Calculator, by formatting data into a matrix, then they tract the ideal point (positives and negatives one) after the normalization process, after that they compute the distance between each alternative and the ideal point, finally the score of each alternative is computed and its used in the distribution flow process.

The TOPSIS method is a multi-attribute decision making technique, it is easy to use however for the vertical handover the performance of TOPSIS is slightly lower in bandwidth and in delay than GRA for interactive and background traffic

### c) Objective function

N mathematical optimization, the term objective function is used to denote a function that serves as a criterion to determine the best solution to an optimization problem. Specifically, it associates a value to an instance of an optimization problem. The goal of the optimization problem is then to minimize or maximize this function until the optimum, by various processes such as the simplex algorithm.

In [12], authors used an objective function technique; input are gathered from different sources user date, network data and policy used. The authors start by collecting information about signal quality, the requested services with the minimal bit rate and the tolerated delay for each candidate RAT. Then, collecting the networks data like delay, bandwidth …etc. after that, policy such as cost, compatibility, trust, preference and capability along with their weights are defined. The weights can be dynamically changed according to the network condition. Finally, with all factors and their weights, the algorithm iterates and computes the best allocation that maximizes the objective function for overall network.

## IV. DISCUSSION

All of the described schemes and methods present merits and inconvenient, in this we will try to give our opinion.

Generally, network-controlled scheme has many problems like, the no taking into account the users requirements since decision is made in the network side, also the starvation problem can appear since the users are not treated with equity because the decision is for the network profit, so this approach present many gaps. The main advantage of the network centric approach is the simplicity and the capacity of the network operator to handle the algorithms.

For the user-controlled scheme, here we have the inverse problems, decisions made in user side present also many problems and big one are the network load balancing since user interest is to allocate resource units to the biggest number of users; this will generate the network loading and saturation. The big advantage is the taking into account the user's preferences of bandwidth and the data rate required for each services. The third kind is the combined one or the collaborative scheme, is the most suitable and appropriate; besides two previously defined approaches, the collaborative approach is the most compromising in terms of profit between users and network operator since it takes into account the profit of both sides for making decisions.

The schemes presented here are mostly mathematical based approaches. The inputs of these functions cover all necessary factors to make a good decision; since user information is so important but implementing schemes will lose transparency, the challenge is finding compromise between utility and cost of involving users.

## V. CONCLUSION

In this paper, a general view of the methods and the schemes used in the network selection problem in a heterogeneous wireless network. We presented all approach with their methods without describing the mathematical modeling; as perspective, we will concentrate our researches for the case of collaborative approach with the two kinds of traffic, the real time as VOIP, video conferencing and the non-real time one like date transfer.